\documentclass[12pt]{article}
\usepackage{hyperref}
\newcommand{\be}[3]{\begin{equation}  \label{#1#2#3}}
\newcommand{\bea}[3]{\begin{eqnarray}  \label{#1#2#3}}
\newcommand{\ee}{\end{equation}}
\newcommand{\eea}{\end{eqnarray}}
\newcommand{\ba}{\begin{array}}
\newcommand{\ea}{\end{array}}
\renewcommand{\arraystretch}{1.4}

\usepackage{amssymb,bbold,euscript}

\begin{document}

\begin{flushright}
\hfill{UPR-1034-T}\\
\hfill{AEI-2003-036}\\
\hfill{hep-th/0303266}

\end{flushright}

\vspace{15pt}

\begin{center}{ \Large
{\bf
Time-dependent backgrounds  from supergravity  \\[3mm]
with gauged non-compact R-symmetry }}

\vspace{20pt}

{Klaus Behrndt$^a$\footnote{E-mail: \tt behrndt@aei.mpg.de}
 and Mirjam Cveti\v c$^b $\footnote{E-mail:
{\tt cvetic@cvetic.hep.upenn.edu}. On sabbatic leave from the
University of Pennsylvania.}
 }
\vspace{15pt}

{\it $^a$ Max-Planck-Institut f\"ur Gravitationsphysik\\
Am M\"uhlenberg 1, 14476 Golm,  Germany\\[4mm]
$^b$ Institute for Advanced Study, Princeton, NJ 08540, USA }
\vspace{30pt}

\underline{ABSTRACT}
\end{center}

We obtain a general class of time-dependent, asymptotically de Sitter
backgrounds which solve the first order bosonic equations that
extremize the action for supergravity with gauged non-compact
$R$-symmetry. These backgrounds correspond only to neutral fields with
the correct sign of kinetic energy. Within $N$=2 five-dimensional
supergravity with vector-superfields we provide examples of
multi-centered charged black holes in asymptotic de Sitter space,
whose spatial part is given by a time-dependent hyper-K\"ahler
space. Reducing these backgrounds to four dimensions yields
asymptotically de Sitter multi-centered charged black hole backgrounds
and we show that they are related to an instanton configuration by
a massive T-duality over time.  Within $N$=2 gauged supergravity in four
(and five)-dimensions with hyper-multiplets there could also be
neutral cosmological backgrounds that are regular and correspond to
the different de Sitter spaces at early and late times.

\newpage



\section{Introduction}


Time-dependent backgrounds in fundamental theory are much less
understood than stationary solutions. In addition these solutions
are typically singular, leading cosmological singularities of the
big bang or big crunch type.

The purpose of this paper is to provide a large class of
cosmological solutions that have an interpretation of solutions of
gauged supergravity with non-compact $R$-symmetry gauged. In this
case the positive cosmological constant leads to asymptotically de
Sitter  space. The obtained class of solutions satisfies first
order bosonic equations that extremize the action, and can be in a
broader sense referred to as supersymmetric (see
\cite{120,510,660} and references \ therein).

Gauging of non-compact $R$-symmetry introduces an effective
imaginary gauge coupling via Wick rotation $g= i\lambda$, and thus
the covariant derivatives for charged fields are typically
complex. In order to have a real Lagrangian, one would have to
impose further Wick rotation for  the charged matter fields or the
gauge fields. In either case, this will change the sign of the
kinetic energy terms for  their respective fields, which in fact
is the case for the supergravity models based on supergroups that
include the de Sitter group \cite{690}.

In this paper we shall consider only the neutral backgrounds that have
the correct signs of the kinetic energy terms.  Nevertheless at the
quantum level one has to face the problem with the ghost-like
contribution from charged sectors.  We do not have much to say about
this problem, and focus only on the neutral, bosonic part, where the
classical backgrounds are turned on.

The time-dependent asymptotically de Sitter backgrounds of
supergravity theory with non-compact $R$-symmetry gauged inherit a
number of properties of the static, asymptotically flat BPS
solutions of ungauged supergravity: there exist multi-centered
solutions, and the flat transverse space can be replaced by a
general hyper-K\"ahler space.

In order to motive the basic set-up, we start with the static
asymptotically flat BPS solution of five- (or four-) dimensional
ungauged supergravity and then generate the corresponding cosmological
solution, that solves the same first order integrability conditions,
but now for (non-compact) gauged supergravity with the {\it positive
cosmological constant}.

As an illustration let us demonstrate the generating technique for
five-dimensional $N$=2 ungauged supergravity without couplings to
matter supermultiplets.  The static BPS black hole solution is given
by
\be010 ds^2 = -{1 \over H^2} dt^2 + H dx^m dx^m \quad , \qquad A =
{dt \over H} \ee
where $H$ is a harmonic function in the transverse
four-dimensional space and $A$ is the $U(1)$ gauge field one-form.
For $H = 1 + {q \over |x|^2}$ this is the extreme
Reissner-Nordstr\"om black hole in an asymptotically flat
Minkowski space.

If one shifts $H$ by a general linear function in the time, i.e.\
\be020 H(x) \rightarrow \lambda \,t + H(x) \ee
the asymptotic space becomes de Sitter and the solution
(\ref{010}) solves the equations coming for the five-dimensional
bosonic action
\be030 S_5 = \int\Big[ {R\over 2} - {1 \over 4} F^2 + {3 \over 2}
\, \lambda^2 \Big] \ . \ee
To verify this, one can calculate the Ricci tensor for the general
function $H= H(r,t)$ and if we separate terms proportional to time
derivatives, the Einstein equations become
\be729
\ba{l}
R_m^{\ n} \rightarrow R_m^{\ n} + [(\partial_0 H)^2
+ {1 \over _2} H \partial_0^2 H ] \, g_m^{\ n} = (F^2)_m^{\ n} -
{1\over 6} \, g_{m}^{\ n} F^2
+ \lambda^2 g_{m}^{\ n} \ , \\
R_0^{\ 0} \rightarrow R_0^{\ 0} + (\partial_0 H)^2 + 2 H
\partial_0^2 H =  (F^2)_0^{\ 0} - {1\over 6} \, F^2
+ \lambda^2 \ , \\
R_0^{\ m} \sim \partial_m\partial_0 H = 0 \ .
\ea\ee
If $\partial_0^2 H = \partial_m\partial_0 H = 0$ the positive
cosmological constant is compensated by the linear time-dependent part
in $H$. Note, the gauge field equations are not effected by the
cosmological constant and with $F_{0m} = {\partial_m H \over H^2}$ one
finds
\be720
\partial_0( \sqrt{g} F^{0m} ) = \partial_0\partial_m H = 0
\quad , \quad
\partial_m( \sqrt{g} F^{m0} ) = \partial_m\partial_m H = 0 \ .
\ee
Thus  the equations of motion are in fact solved  with  the Ansatz
(\ref{020}) and the solution becomes after time reparameterization
equivalent to the ones found in \cite{600}.

If one turns off the $U(1)$ gauge field charge, $H(x) = 1$, and
the space-time metric becomes
\be926 ds^2 = - {dt^2 \over (\lambda \, t + 1)^2 } + (\lambda \, t
+ 1) dx^m dx^m \ee
which corresponds to the five-dimensional de Sitter space.

We shall see that the harmonic function Ansatz of the type
(\ref{020}) solves the equations of motion even for more general
Lagrangians, that involve more $U(1)$ gauge fields and the scalar
fields.  In a more general set-up, we focus on the prototype
example of five-dimensional $N$=2 gauged supergravity with
vector-superfields. We will also allow for the initial
configuration to be stationary. In Section 2 we discuss
multi-centered, multiple charged Abelian black holes with
asymptotic de Sitter space, which have been found in \cite{590} as
solutions of first order equations that extremize the action.

In Subsection 3.1 we further reduce these solutions to
four-dimensions, by first replacing the spatial part of the five
dimensional solution with a specific time-dependent hyper-K\"ahler
space with $U(1)$ isometry. Reduction of these backgrounds to four
dimensions yields four-dimensional, asymptotically de Sitter
multi-centered charged black hole backgrounds.

By turning off the gauge fields we further discuss the backgrounds
with only scalar fields coupled to gravity in Subsection 3.2.
These backgrounds correspond to the cosmological flows, that are
complementary to the BPS domain walls for supergravity theories
with gauged compact symmetry \footnote{BPS domain
wall solutions of supergravity theory we first found in the
context of $N$=1 D=4 ungauged supergravity in \cite{CGR}.  (See
also \cite{CGS} for generalizations and \cite{CS} for a review.)},
i.e. renormalization group (RG) flows \cite{620} in the context of
AdS/CFT correspondence \cite{760}.

In Subsection 3.3 we also address the possible resolutions of the
early time (cosmological) singularity that are generic in the set
up with vector-supermultiplets.  Within $N$=2 gauged supergravity
coupled to hyper-multiplets, one can also obtain regular
cosmological backgrounds with a different value of the de Sitter
cosmological constant at early and late times and we comment on
this possibility.

Examples of four-dimensional charged de Sitter black hole
backgrounds can be obtained by performing (massive) T-duality over
the time if one starts with the  BPS instanton solution of
ungauged supergravity.  We discuss the procedure in Section 4.  We
also note that the obtained five- and four-dimensional gauged
supergravity backgrounds should be related to a specific reduction
of  D=10 type II$^{\star}$ theories \cite{120} on non-compact
spaces.

Conclusions and open questions are addressed in Section 5.


\section{Cosmological background in five dimensions}


In the Introduction we demonstrated that in the case of static BPS
solutions of $N$=2 ungauged supergravity without couplings to
matter multiplets, time-dependent backgrounds of gauged
supergravity with a positive cosmological constant can be obtained
by replacing the integration constants in the harmonic functions
by a function linear in time (\ref{020}). Via this procedure
static, asymptotically flat BPS solutions of ungauged supergravity
become time-dependent, asymptotically de Sitter solutions of
gauged supergravity (with non-compact $R$-symmetry gauged).

In this section we will show that such time-dependent backgrounds,
in a general set-up, solve first order differential equations
which extremize the action. As a prototype example we shall focus
on the five-dimensional $N$=2 gauged supergravity which couples to
vector-supermultiplets, only.

The gravity supermultiplet has, besides the graviton and
gravitino, one Abelian vector field and each vector-supermultiplet
has one Abelian vector field and a (real) scalar field $\phi^A$
($A=1, ... , n_v$) that parameterize a manifold ${\cal M}$ defined
by the constraint
\be772
{\cal V} = {1 \over 6} C_{IJK} X^I X^J X^K = 1
\ee
with $I= 0,1, ..., n_v$. The constants $C_{IJK}$ enter the
Chern-Simons term of the action
\be031
S_5 = \int\Big[ {R\over 2} - g^2 V - {1 \over 2} g_{AB}
\partial \phi^A \partial \phi^B - {1 \over 4} G_{IJ} F^I \cdot F^J
\Big] + {1 \over 12} \int C_{IJK} F^I \wedge F^J \wedge A^K
\ee
and $F^I$ are the field strength for the $U(1)$ gauge fields and
the potential reads
\be823
V = 6 \,  \Big( {3 \over 4} g^{AB} \partial_A W \partial_B W - W^2 \Big) \ .
\ee
The couplings in the Lagrangians are now defined by
\be652
G_{IJ} = {1 \over 2} \partial_I \partial_J {\cal V} \Big|_{{\cal V}=1}
\quad , \qquad g_{AB} = \partial_A X^I \partial_B X^J G_{IJ}
\ee
where $\partial_A X^I \equiv {d \over d\phi^A}   X^I(\phi^A)$.
With this definitions one finds the relations
\be251
\ba{l}
X_I = {2 \over 3} G_{IJ} X^J \quad , \qquad dX_I = - {2 \over 3} G_{IJ}
dX^J  \ , \\
{\cal V} = X^I X_I = 1 \quad , \qquad X_I d X^I = X^I d X_I = 0  \ .
\ea
\ee
Unbroken supersymmetry implies the existence of at least one Killing
spinor $\epsilon$ which gives a zero for the gravitino and gaugino
supersymmetry variation. If one includes an Abelian gauging of the
$SU(2)$-$R$-symmetry with the gauge field $A = \alpha_I A^I$ (with
$\alpha_I=const.$), these variations contain a superpotential of the
form \cite{160}
\be722
W= \alpha_I X^I \ .
\ee
With the definition of the scalar field metric, we can write
$g_{AB} d \phi^B = -{3 \over 2} \partial_A X^I dX_I$ and $\partial_A W
= \partial_A X^I \alpha_I$ and the supersymmetry variations become
\be625
\ba{rcl}
\delta \Psi_\mu &=& \Big[D_\mu +
        {i \over 8} \Big( \Gamma_{\mu}^{\ \ \alpha\beta}
        - 4 \delta_{\mu}^{\ \alpha} \Gamma^\beta \Big) X_I F_{\alpha\beta}^I
        + {1 \over 2}  g \, \Gamma_{\mu}  \, W
        \Big] \epsilon = 0 \ , \\
\delta \lambda_{A} &=& {3i \over 2}  \partial_A X^I
\Big[ \Gamma^\mu \partial_\mu
X_I + {2i \over 3} G_{IJ} F^J_{\mu\nu} \Gamma^{\mu\nu} + {1 \over 2}
g \alpha_I \Big] \, \epsilon = 0 \ , \\
D_\mu &=& \partial_{\mu} +{1 \over 4} \omega_\mu^{ab}
\Gamma_{ab} - {3i \over 2} g \alpha_I A^I_{\mu} \ .
\ea
\ee
To keep the notation as simple as possible we have not used symplectic
Majorana spinors but used instead the conventions of \cite{160}.
Solutions of these first order differential equation solve
the equations of motion if in addition the gauge field equations
are satisfied, i.e.
\be628
dF^I = 0 \quad , \qquad
d \, \big( \, G_{IJ} {^\star F}^J + {1 \over 2}
C_{IJK} A^J \wedge F^K \, \big) =0 \ .
\ee

Let us recall first the supersymmetric solution in the ungauged case
($g=0$).  It has a time-like isometry and can be written in a proper
coordinate system as \cite{150,130}
\be100
\ba{rcl}
ds^2 &=& - e^{- 4U} (dt + \omega)^2 + e^{2U} dx^m dx^m \ , \\
A^I &=& e^{-2U} X^I (dt + \omega) \ , \\
e^{2U} X_I &=& {1 \over 3} H_I  \quad , \qquad
d \omega + {^\star d\omega} = 0 \ .
\ea
\ee
The one-form $\omega = \omega_m dx^m$ corresponds to $U(1)$ fibration of
the transversal space and without specifying the functions $H_I$,
these fields solve the equations (\ref{625}). However, in order to fulfill
the equations of motion we have to consider the gauge field equations
(\ref{628}) that are solved only if
\be727
\partial_m \partial^m H_I \ .
\ee
Recall, the fields  $X^I$ are subject to the constraint (\ref{772})
so the $n_v + 1$ harmonic functions determine the scalar fields
$\phi^A$ as well as the metric function $e^{2U}$. Note, $g=0$
corresponds to the case of ungauged supergravity.

If we consider a  gauging of  non-compact  $R$-symmetry, i.e. $g^2 <
0$, the equations of motion  arising from the action (\ref{031})
are solved if
\be273
(- \partial_0^2 + \partial_m\partial_m) H_I = 0
\quad , \qquad \partial_0 H_I = \alpha_I \lambda \quad ,\qquad
g = i \lambda
\ee
are i.e.\ $H_I$ is now harmonic in all five coordinates and the
vacuum, given by stationary point of the superpotential ($dW =0$),
yield now a de Sitter space (since $g^2 <0$).  Note that we are
tacitly assuming that the first order equations arising from the
fermionic supersymmetry variations (\ref{625}) are formally still
valid, in spite of the fact that $g=i\lambda$ is an imaginary
coupling.

The proof that the equations of motion are solved with the above
Ansatz for the modified harmonic functions has already been given in
\cite{590} (partly based on \cite{600}).  In our context we use a
different time parameterization where $H_I$ depend linearly on
time. We have summarized the proof that the first order equations
(\ref{625}) are in fact solved in the Appendix.

In the following we shall discuss only the gauge field equations
(\ref{628}), which yield harmonic functions as a solution.  Using the
form for the gauge field strength (\ref{153}), the Chern-Simon term
can be written as
\be799
{1 \over 2} C_{IJK} F^J \wedge F^K =
d \Big[ e^{-6U} H_I (dt + \omega) \wedge d\omega \Big]
\ee
and moreover, using the explicit form (\ref{666}) (and $\omega_m
\partial_m H_I = \omega_m \partial_m U = 0$) we find
\bea776
\partial_n \Big( G_{IJ} \sqrt{g} F^{I\, n0}\Big)
&=& \partial_n \partial_n H_I - \partial_n \Big( e^{-6U} H_I \omega_m
\partial_{[n} \omega_{m]} \Big) \ , \\
\partial_\mu \Big( G_{IJ} \sqrt{g} F^{I\, \mu n }\Big) &=&
\partial_0 (\omega_n \partial_0 H_I ) - \partial_0 \partial_n H_I
- \partial_\mu \Big( e^{-6U} H_I \omega_m
\partial_{[n} \omega_{m]} \Big)
\eea
where $ \mu = 0,1,2,3,4$. For trivial $\omega$, these equations agree
with (\ref{720}) and are obviously solved, but if $\omega$ is non-trivial
we have to use the fact that $d\omega$ is anti-selfdual in order to
verify that the equations are solved if (\ref{273}) holds.

It is obvious from eq.\ (\ref{273}) that the equations of motion
are also satisfied for the multi-centered solution, and that the
flat transverse space can be replaced by a hyper-K\"ahler space
(see the next Section). Therefore, these asymptotically de Sitter
solutions inherit features of the asymptotically flat BPS
solutions of ungauged supergravity.


\section{Cosmological background in four dimensions}


The procedure, described in the previous Section to obtain
time-dependent backgrounds from BPS solutions can of course also
be applied to other dimensions. Again, one has to shift the
harmonic functions by functions linear in time, combined with the
replacement  $g = i \lambda$. Again, we consider a truncation to
the neutral sector only where all kinetic terms have the correct
sign.

We shall however not give a general proof of the procedure, but will
focus instead on deriving the four-dimensional cosmological background
by dimensional reduction of the five-dimensional one described in the
previous Section.


\subsection{Transversal hyper-K\"ahler space and the reduction to
four dimensions}


Following the calculations in the Appendix, it is straightforward
to realize that one can replace the flat transversal space of the
five-dimensional metric in (\ref{100}) by any four-dimensional
hyper-K\"ahler space. To see this, let us write the metric as
\be662
ds^2 = - e^{- 4U} (dt + \omega)^2 + e^{2U} h_{mn} dx^m dx^n  \ .
\ee
This yield instead of (\ref{552}) the equation
\be770
[\nabla_\mu^{(h)} + (\partial_\mu U) ]\epsilon = 0 \ .
\ee
with the covariant derivative corresponding to the metric
\be444
ds^2 = - dt^2 + h_{mn} dx^m dx^n \ .
\ee
As a solution one finds that $h_{mn}$ can be any Ricci-flat
metric. On any constant time slice, the integrability condition
for this equation means that the Riemann tensor has to be
self-dual and hence the space is hyper-K\"ahler (and therefore
Ricci flat).

For concreteness we consider the four-dimensional Gibbons-Hawking
metric \cite{201}
\be700
h_{mn} dx^m dx^n = {1 \over V} (d\chi + \sigma )^2 + V dx^i dx^i
\quad , \qquad \partial_i V = \epsilon_{ijk} \partial_j \sigma_k
\ee
with the isometry $\partial_\chi$. The equation for $V$ is solved with
any harmonic function in the three-dimensional space spanned by the
coordinates $x^i$. We are however interested in a time-dependent
functions, i.e.\ $V=V(t,x)$.  Inserting in (\ref{444}) one
can verify by an explicit calculation that the five-dimensional space
remains Ricci-flat as long as
\be920
\partial_0 \partial_0 V = \partial_0 \partial_i V = 0 \ .
\ee
Thus, this harmonic function is on equal footing to the harmonic
functions introduced in (\ref{273}).

Assuming that $\partial_\chi$ is also an isometry of the complete
five-dimensional metric, it is straightforward to reduce the model to
four dimensions.  In order to obtain a diagonal metric in four
dimensions, which may be most interesting from the cosmological point
of view, we identify the $U(1)$ fibre $(d\chi + \sigma)$ with $\omega$
in the five-dimensional metric (see also \cite{100}), which yield the
four-dimensional metric
\be651
ds^2 = - e^{-2\hat U} dt^2 + e^{2 \hat U} dx^i dx^i \ .
\ee
In order to obtain the Einstein-frame metric (\ref{651}) after the
reduction,  there is a conformal rescaling of the metric,  and the
function $\hat U$ is given by
\be788
e^{4 \hat U} = V e^{6U} \ .
\ee
We expect, that the general four-dimensional solution can also be
obtained directly as  a generalization of the general BPS solution
\cite{210}, where the complex scalar fields $z^A = {Y^A \over
Y^0}$ and the metric function $e^{2\hat U}$ are now a solution of
the equations
\be529
\ba{l}
e^{2 \hat U} = i (\bar Y^I F_I - Y^I \bar F_I)
\quad , \\
i(Y^I - \bar Y^I) = H^I(t,x)  \quad , \qquad
i(F_I - \bar F_I) = H_I(t,x)
\ea
\ee
with: $(H^I(t,x) = \lambda \, \beta^I \, t + h^I(x)$ and $H_I(t,x)
) = \lambda \, \alpha_I \, t + h_I(x))$. As in five dimensions,
these equations should solve the BPS equations of four-dimensional
supergravity with a gauged $R$-symmetry (and with $g=i\lambda$).
Note, in this solution the symplectic section $(Y^I, F_I)$ is
rescaled so that it becomes invariant under K\"ahler
transformations (see \cite{390, 210} for details).  A number of
explicit solutions of these algebraic equations have been
discussed in the literature \cite{290}. [Note, $(h^I(x), h_I(x))$
is again a set of any harmonic functions and includes especially
also multi-center black holes.]  We will show in the next Section
that the shift by a linear function in time is the result of a
massive T-duality over time applied on an instanton solution of
ungauged supergravity. The massive T-duality generates in the dual
theory the correct potential.

We conclude this Subsection by presenting a specific example where
the Chern-Simons cubic form (\ref{772}) reads
\be655
{\cal V} = X^1 X^2 X^3 \ .
\ee
This is the so-called STU model ($S=X^1, T=X^2 , U=X^3$) with
three Abelian gauge fields and two real scalars, e.g.,\ by $\phi^1
= X^2/X^1$ and $\phi^2 = X^3/X^1$. For the single center case, one
obtains for the five-dimensional solution reads
\bea320
e^{6U} &=&  H_1 H_2 H_3 = \Big(\lambda \,t + {q^1 \over r^2}\Big)
\Big(\lambda \,t + {q^2 \over r^2}\Big)
\Big(\lambda \,t + {q^3 \over r^2}\Big) \quad , \qquad \omega = 0 \ , \\[2mm]
X^I &=& {e^{2U} \over H_I}
\quad  , \qquad
A^I = {dt \over H_I } \ .
\eea
In order to reduce this model to four dimensions, we will replace
the transversal space by a specific hyper-K\"ahler space given by
the Gibbons-Hawking metric.  This requires, that the harmonic
function in $e^{6U}$ do not depend on $\chi$, which has to be a
isometric direction.  Hence, we have to replace the harmonic
functions: $\big(\lambda \, t + {q^i \over r^2}\big) \rightarrow
\big( \lambda \, t +{q^i \over r}\big)$, where $r$ is now the
three-dimensional radial direction ($dx^idx^i \equiv dr^2 + r^2 d
\Omega_2$).  Denoting $V = \lambda t + {q^0 \over r}$ the
four-dimensional metric function becomes
\be605
e^{4 \hat U} = \Big(\lambda t + {q^0 \over r}\Big)
\Big(\lambda \,t + {q^1 \over r}\Big)
\Big(\lambda \,t + {q^2 \over r}\Big)
\Big(\lambda \,t + {q^3 \over r}\Big) \ .
\ee
Of course we can replace ${q^m \over r}$ by any harmonic function
giving a general multi-center solution living in an asymptotic de
Sitter space. This is the de Sitter analog of the four-charge
black hole of ungauged supergravity \cite{790}.  In fact, for
large $t$ the metric (\ref{651}) becomes
\be002
ds^2 = -{dt^2 \over (\lambda \, t)^2} + (\lambda \, t)^2 dx^i dx^i \ .
\ee
On the other hand, if we go back in time, we will necessarily reach
the point where $e^{4 \hat U}$ becomes zero giving a curvature
singularity, i.e.\ a big bang singularity.


\subsection{Cosmological renormalization group flow}


If we turn off the $U(1)$ gauge fields, the background becomes only
time-dependent (any dependence on the spatial coordinates disappears)
and we can interpret the time evolution as a cosmological
renormalization group (RG) flows within in the proposed dS/CFT
correspondence \cite{720}. Let us formulate the flow equations in $D$
dimension and write the (Robertson-Walker) metric as
\be210 ds^2 = -d\tau^2 + e^{2A(\tau)} \, dx^i dx^i \ . \ee
 If we
write the Lagrangian as
\be542
S_D = \int \Big[ {R \over 2} - g^2 V - {1 \over 2} g_{AB} \partial \phi^A
\partial \phi^B \Big]
\ee
where the potential is given in terms of a real superpotential $\tilde W$
\be728
V = {(D-2)^2 \over 2} \partial_A \tilde W \partial^A \tilde W -
    {(D-1)(D-2) \over 2} \, \tilde W^2 \ .
\ee
In five dimensions (\ref{542}) corresponds to Lagrangian (\ref{031})
without $U(1)$ gauge fields.
In four dimensions the $\tilde W$ is related to the complex
superpotential by $\tilde W^2 = e^{K} W \bar W$.  Because $g^2 = -
\lambda^2$ the potential has the opposite sign and we can square the
action for the time-dependent metric (\ref{210}) in the same way as
for BPS domain walls (see also \cite{800,CGR}) and we find
\be230
\ba{l}
S = S_{bulk} + S_{boundary} \\
S_{bulk} = \int e^{(D-1)A} \Big[-(D-1)(D-2) (\dot A - \lambda \tilde W )^2
    +{1 \over 2} |\partial \phi^A + (D-2)\, \lambda \, g^{AB}
    \partial_B \tilde W|^2 \Big]\ .
\ea
\ee
{From} here we obtain the first order equations
\be240
\dot A = \tilde W
\quad , \qquad
\dot \phi^A = - (D-2) \, g^{AB} \partial_B \tilde W \ .
\ee
Extrema of $\tilde W$ correspond obviously to dS vacua and the
solution fulfilling these first order flow equations  extremizes
the action and  de Sitter vacua are approached with vanishing
velocity  with no  oscillations.

These first order equations corresponds to the cosmological
RG-flow equations. Regular flow requires necessarily a
superpotential $\tilde W$ with at least two (connected) extrema.
For the examples at hand (i.e.\ vector-multiplets, only), one can
however show \cite{670} that on any given component of the scalar
manifold, there is at most one critical point and therefore a
given solution will always run towards a singularity related to a
pole in $\tilde W$. In the cosmological setting this means that
the big bang singularity is un-avoidable in these models.  However
one should of course ask, whether one can trust the model all the
way to the singularity or whether one should replace the model by
a different one -- as it has been discussed for the cusp
singularity for ordinary AdS domain walls.


\subsection{Resolution of the singularities}


There are different ways in dealing with a singularity, e.g.:
\medskip

$(i)$ it may be an artifact of the approximation or truncation of the theory,

$(ii)$ it appears in a non-physical region, which should be replaced or

$(iii)$ the singularity could be smoothed out by a second de
Sitter space.

\medskip

\noindent
Let us comment on either case.

Case $(i)$: It is well known that the appearance of singularities
in supergravity reflects often the failure of the approximation.
However, in most interesting examples one is forced to consider
certain approximations as e.g.,\ the restriction to the lowest
order in the expansions in $\alpha'$ and/or $g_s$. It is likely
that higher derivative corrections coming from the $\alpha'$
expansion can smooth out singularities, e.g.,  as  it is known in
Born-Infeld instead of Maxwell theory. Another source of
singularities is the truncation to Abelian gauge fields instead of
non-Abelian, which yields a regular background,  such as an
example  discussed in \cite{580}.

Another interesting way to regulate supergravity solutions is via the
so-called transgression mechanism \cite{180}, where the incorporation
of additional differential forms yield a regular background due to
corrections coming from the Chern-Simons term.  For five-dimensional
supergravity, this mechanism was in fact used in \cite{100} to obtain
regular BPS solutions in ungauged supergravity. In our derivation it
was important that $e^{-2U} d\omega$ as well as the spatial components
of the gauge fields are anti-selfdual, which yielded the harmonic
equations for $H^I$. In the case with no time-dependence
($\lambda=0$), it was shown in \cite{100} that one can add a self-dual
2-form $G^{(+)}$ to $e^{-2U} d\omega$ as well as to the spatial
components of the gauge field strength. As a result, the Chern-Simons
transgression term gives an additional contribution from $G^{(+)}
\wedge G^{(+)} = |G^{(+)}|^2$ and the gauge field equations do not
yield the harmonic equation for $H_I$, but: $\partial_m \partial_m H
\sim |G^{(+)}|^2$. This modification via the transgression term
resolves singularities \cite{180} related to poles in $H$.  However,
for $\lambda \neq 0$ the cosmological singularity appears if one goes
back in time, i.e.\ $H$ decreases and may eventually vanish. At this
point however, the time-derivatives of $H$ are still
smooth and hence, the transgression mechanism cannot resolve these
singularities. Note, that even for the pole singularities, this
mechanism seems to fail if one includes the time-dependence ($\lambda
\neq 0 $), this however deserves further investigations.

Case $(ii)$. The cosmological singularity of the discussed
solutions is reminiscent of the repulson singularity discussed for
black holes \cite{300}.  This singularity appears in the
supergravity solution as a zero of the metric function $e^{2U}$
(or $e^{2 \hat U}$ in four dimensions) due to a vanishing or
negative harmonic function. Since the harmonic functions
parameterize the radii of the internal cycles, this singularity
indicates interesting physics related to gauge symmetry
enhancement, flop or conifold transitions. By using a probe
analysis for time independent black holes one can show, that even
before one reaches the repulson singularity the tension of the
probe brane becomes tensionless and at the enhancement point the
supergravity solution cannot be trusted anymore and additional
massless degrees of freedom have to be taken into account
\cite{340}. An analogous mechanism does also apply for (static)
domain walls, where the enhancon locus appears before one reaches
the space-time singularity \cite{370} and integrating-in the
additional modes, the singularity can be avoided (see \cite{350}
and references therein).

We would like to argue that also for the cosmological solutions at
hand the repulson singularity is only an artifact of the
approximation and can be smoothed out. For the four-dimensional
case discussed in Subsection 3.1 there is a powerful tool to
regulate supergravity solutions, by including the term
proportional to the Euler number in the prepotential. This is an
$\alpha'$-correction which is added to the prepotential as
\be773
F = -i \, {\chi \zeta(3) \over 2 (2\pi)^3} (Y^0)^2 + \tilde F(Y)
= -i (Y^0)^2 \Big[ c  + {\cal F}(z) \Big]
\ee
with $c= {\chi \zeta(3) \over 2 (2\pi)^3}$ and $\chi$ is the Euler
number of the internal space.  Recall, we use the rescaled section
$(F_I(Y) , Y^I)$, which ensures that the supergravity solution is
invariant under K\"ahler transformations (by rescaling of the section
with the supersymmetry central charge \cite{390}). In this rescaled
section we should not set $Y^0 = 1$ because this would eliminate one
harmonic function in (\ref{529}). Now, if the scalars flow to smaller
values and even if one reaches the point where ${\cal F}(z)$
vanishes, the metric will still be regular as long as $c
\sim \chi \neq 0$ (see also \cite{290}). In fact, if we solve the
equations (\ref{529}) for this prepotential the metric function
becomes
\be771
e^{2 \hat U} \sim H_0^2  \qquad {\rm for} \quad {\cal F}(z)  \simeq 0 \ .
\ee
and we obtain a de Sitter space if we choose $H_0 = \lambda (t
-t_0)$. [In the context of transgression mechanism for the static
BPS configurations the  $\alpha'$ correction due to the
gravitational Chern-Simons term, proportional to the Euler number
of the internal space,  was studied in \cite{BCN}.]

Another form of singularity is of course given by scalars that run
to infinite values, which do not correspond to divergencies of the
internal space but correspond to a large volume limit.  In this
case, the singularity reflects  the appearance of an extra
dimension and the corresponding higher dimensional solution should
be regular. We have not much to say about this singularity (see
\cite{480} for a recent discussion), but this remark also brings
us to a discussion of the situation in five dimensions. There, the
supergravity solution describes also a flow with a de Sitter
vacuum at late time and may run towards a phase transition point.
In five dimensions however, we do not have the Euler number term
in the pre-potential and hence the mechanism as described in four
dimensions is not applicable. Nevertheless, the solution is
regular as long as at least some cycles stay finite as, e.g., at
the flop transition point \cite{490,570}, but also at boundaries
of the vector moduli space.  Note, that although scalars disappear
at this point, other scalars may continue to flow along this
boundary to infinity, indicating the appearance of additional
large extra dimensions.

Case $(iii)$. The resolution that we discussed so far, rely on the
picture that scalar fields parameterize cycles of an internal
space and the model may change if the scalars vanish. {From} the
supergravity point of view however, all these flows are singular.
To regularize the flow, one has to consider a superpotential with
a second extremum. Complementary to BPS domain walls, this flow
would interpolate between two anti de Sitter vacua; in this case
we would have a cosmological flow connecting an early de Sitter
phase with a late time de Sitter space. In fact, if one includes
hyper-supermultiplets it is possible to construct smooth flow
\cite{640}, which represents a truncation of the flow found in
\cite{620} on $N$=2 supergravity. However, we have to truncate the
model onto the neutral scalar sector and one should investigate if
this can be done consistently. Similarly, it would be interesting
to see whether the superpotential yielding a realization of the
Randall-Sundrum model in supergravity \cite{630} can be employed
to obtain a regular time-dependent solution.  Since in this case
the warp factor is exponentially small at both sides of the wall,
it would describe a bubble that first exponentially expands
followed by an collapse.


\section{Time-like ``massive'' T-duality in D=4}


We will now show that the cosmological solution described in Section 3
can be obtained from a BPS configuration of ungauged supergravity by a
T-duality along the time. It is however not the usual T-duality
transformation, but the ``massive'' one, where one takes into account
a linear dependence with respect to the isometry direction
\cite{250}. For concreteness we will use a ``massive'' generalization
of the c-map, which maps type IIA and IIB superstring compactification
on the same internal space onto each other and which is nothing but a
time-like T-duality \cite{410,420}.

To explain the procedure we start with the simple case, where we
have only the gravity supermultiplet, but no
vector-supermultiplets and no potential.  In this case, the BPS
solution is just the Reissner-Nordstr\"om black-hole (with or
without rotation). As a result from the c-map, the two on-shell
degrees of freedom of the metric as well as of the two degrees of
freedom of the gauge field combine to four scalars of the
universal hyper-supermultiplet. Therefore, in type II
compactifications two scalars are in the NS-NS sector and two
coming from the RR sector and the BPS configuration becomes an
instanton solution with a flat Einstein frame metric (see
\cite{430}). The scalars of the classical universal
hyper-supermultiplet spans a manifold given by the coset ${SU(2,1)
\over U(2)}$ and can be decomposed into two complex scalars. For a
BPS configuration, one complex scalar is trivial (constant) and
the other is fixed by the BPS equations and parameterizes the
coset ${SU(1,1) \over U(1)}$ as given by the sigma model
\be723
{\partial S \partial \bar S \over ({\rm Im} S)^2} \ .
\ee
There are now two different possibilities to embed this coset into
${SU(2,1) \over U(2)}$. In one case one combines the two NS-NS-scalars
into the complex field
\be902
S= a + i e^{-2\varphi}
\ee
which is the well-known combination in heterotic string models, where
the axion $a$ is dual to the anti-symmetric tensor field and $\varphi$
is the dilaton. The other possibility is to combine one NS-NS
and one RR-scalar yielding
\be901
S= a + i e^{-\varphi}
\ee
which is the combination appearing in type IIB compactifications.
This is also the representation that we obtain after the c-map of the
type IIA model.

As the next step, we show that a solution of the instanton equation
can be mapped onto cosmological solutions as given in eqs.\
(\ref{651}) with $e^{2 \hat U}$ as a harmonic function. Following
\cite{440} we are looking for an instanton solution in Euclidean time
with a vanishing energy-momentum tensor. Due to the Wick rotation the
sign of the axionic part in the Lagrangian is changed
\be659
{\partial S \partial \bar S \over
({\rm Im} S)^2}
=(\partial \varphi)^2 + e^{2 \varphi} (\partial a)^2
\quad  \longrightarrow \quad
{\partial S_+ \partial  S_- \over
{1 \over 2} (S_+ - S_-)^2} =
 (\partial \varphi)^2 - e^{2 \varphi} (\partial a)^2
\ee
with $S_\pm =e^{-\varphi} \pm a$. For the instanton configuration the
energy momentum tensor has to vanish, implying $\partial a = \pm
\partial e^{-\varphi}$ and the equations of motion for the axion
becomes
\be661
\partial_\mu \big( e^{2 \varphi} \partial^\mu a \big)
= \pm \partial_\mu \partial^\mu e^{\varphi} = 0
\ee
and therefore the solution is expressed in terms of a harmonic
function $H$
\be701
e^{\varphi} = H \quad , \qquad a = \pm {1 \over H}+ const.
\ee
In general the harmonic function can depend on all four Euclidean
coordinates, but for the case at hand we do not consider a dependence
on the Euclidean time.

If one is doing the standard T-duality along the Euclidean time
followed by a Wick rotation to the Minkowskean time one obtains the
static Reissner-Nordstr\"om black hole \cite{430}. In order to
generate a time-dependence we will employ the massive T-duality,
which, as the usual T-duality, defines a map of two models
dimensionally reduced over inverse radii, where one model is massless
(i.e.\ no potential) and the other is massive (i.e.\ with a
potential). To make the map explicit one has to use the Scherk-Schwarz
reduction \cite{450}.  This is a generalized dimensional reduction
where one allows for a linear dependence on the coordinate along which
one makes the reduction, but in a way that the reduced model is still
independent of this coordinate.  As a consequence one obtains the
correct potential of massive supergravity, at least as long as one is
doing the T-duality along the spatial direction \cite{110,250,460}.
For the above instanton solution this implies, that we have to perform
the usual dimensional reduction along the Euclidean time followed by
an inverse Scherk-Schwarz reduction over the inverse radius.  The
correct time dependence can be fixed using the global
symmetries. Namely, the time independence of the reduced model is
ensured if it can be absorbed into a $SL(2,R)$ symmetry transformation
of the scalar matrix of the form \cite{460}
\renewcommand{\arraystretch}{1.0}
\be119
{\cal M} = e^{\varphi} \left(\ba{cc} |S|^2 & a \\ a & 1 \ea \right)
 \ \longrightarrow \ \Omega^{-1}(t) \, {\cal M} \, \Omega(t)
\quad , \qquad \Omega^{-1}(t) \, \partial_0 \Omega(t) = C
\ee
where $\Omega \in SL(2,R)$ and the mass matrix $C$ has to be time
independent, yielding the potential \cite{460}
\be502
V \sim {\rm tr} \big( C^2 + C^T {\cal M} C {\cal M}^{-1} \big) \ .
\ee
Next, the harmonic function that defines the solution in (\ref{701})
can be shifted by a constant $H \rightarrow H + h$ by the following
$SL(2,R)$ transformation
\be022
\Omega =\Omega_1^{-1} \, \Omega_2 \, \Omega_1
=  \left(\ba{cc} 1 & 0 \\ {h \over 2} & 1 \ea \right)
\ee
with the two generators $\Omega_1: \ S_\pm \rightarrow S_\pm^{-1}$
and $\Omega_2:\ S_\pm \rightarrow S_\pm + {h \over 2}$ (see
\cite{470} where this transformation was used in type IIB string
theory). In order to fix the time-dependence of $\Omega$ we have
to ensure that the mass matrix $C$ does not depend on time, which
is the case for $h = g \, t$. So, as result of the ``massive''
T-duality we added to $H$ a linear function in time and the
potential (\ref{502}) reads
\be609
V \sim g^2 e^{2 \varphi} \ .
\ee
Finally, to obtain a Minkowski signature one has to perform a Wick
rotation in the time ($t \rightarrow it$) and in $g =i\lambda$, so
that $H$ remains real, which in turn changes the sign of the
potential. Applying standard rules of T-duality, one finds that
the new dilaton is constant and $e^{2\hat U}=e^{2 \varphi}$ and
hence the factor $e^{2 \varphi}$ in the potential is nothing but
$\sqrt{\det(g_{mn})}$ for the Reissner-Nordstr\"om black hole (see
eq.\ (\ref{651})). Therefore  there is  only a positive
cosmological constant in the dual Lagrangian.

The procedure that we described here for the supergravity
supermultiplet can be generalized to include any number of
vector-supermultiplets, which for ungauged supergravity was done
in \cite{430, 520}. In this case one obtains $n_V +1$
hyper-supermultiplet with the scalar fields: $ S, z^A , \xi^I ,
\tilde \xi_I$ (recall: $I = 0,1, ... , n_V$, $A = 1, ... , n_V$).
The explicit transformation $\Omega$ is now much more involved,
but it is again basically given by  shifts of the axionic scalars
$(\xi^I , \tilde \xi_I )$, which represent isometries of the
scalar manifold and which are related to harmonic functions (see
also \cite{520}). However, we do not need the explicit form of
$\Omega$; the correct form of the potential can also be derived if
one replaces the axionic scalars as follows
\be882
(\tilde \xi_I , \xi^I)\  \rightarrow \ (\tilde \xi_I + g \alpha_I t\; , \;
\xi^I + g \beta^I t)
\ee
which for the single axion $a$ in (\ref{659}) yields exactly the
potential (\ref{609}). Note, that up to a U-duality transformation
(inversion of the dilaton field) the transformation $\Omega$
corresponds precisely to a linear time shift in the axion. Using
the notation from \cite{530} the corresponding part of the
hyper-supermultiplet Lagrangian reads
\be092
e^{2 \varphi} (\partial \tilde \xi_I + {\cal N}_{IL} \partial \xi^L)
({\rm Im} {\cal N}^{-1})^{IJ}
 (\partial \tilde \xi_J + \bar {\cal N}_{JK} \partial \xi^K)
\ee
where the complex matrix ${\cal N}$ defines the gauge field
couplings in the Lagrangian. This yields, after the time shifts
(\ref{882}), the potential
\be822
V \sim g^2 e^{2 \varphi} (\alpha_I + {\cal N}_{IL} \beta^L) ({\rm Im} {\cal
N}^{-1})^{IJ}(\alpha_J + \bar {\cal N}_{JK} \beta^K)
\ee
which can be written in the standard potential form appearing in
gauged supergravity \cite{530}. As before, after the time-like
T-duality one has to do a Wick rotation in time, combined with the
replacement $g = i \lambda$, which changes the sign of the
potential resulting in a de Sitter instead of anti de Sitter
vacuum.

In addition to the shifts in the axionic scalars $(\tilde \xi^I ,
\xi_I)$ one could also allow for shifts in the axion $S + \bar S$,
but as result, the dilaton  now appears in the potential and
hence, as a result of the c-map one obtains a model that contains
also a  universal hyper-supermultiplet. It is of course
interesting to explore this possibility further.

In summary, we have seen that a time-dependent background can be
obtained by a time-like (massive) T-duality, which includes a
Scherk-Schwarz reduction over time.  In comparison to massive
T-duality over a space-like direction, the overall sign of the
potential has changed, but otherwise it is the same potential.  Hence,
it is similar to the models introduced by Hull \cite{120}.


\section{Discussion}


In this paper we presented a procedure to generate time-dependent
backgrounds starting from stationary BPS solutions of ungauged
supergravity. We focused on examples of $N$=2 supergravity coupled
to vector-supermultiplets.  These solutions are given by a set of
harmonic functions that can be shifted by linear functions in time
and solve the first order differential equations which coincides
with the BPS equations of gauged supergravity, but with an
imaginary gauge coupling $g=i\lambda$. Since only the $R$-symmetry
is gauged, an imaginary $g$ reflects  the non-compactness of the
$R$-symmetry. Since the bosonic fields are not charged under the
$R$-symmetry, the bosonic model is well-defined. All kinetic terms
in the bosonic Lagrangian have the correct sign, but the potential
has the opposite sign yielding stable de Sitter vacua, instead of
anti de Sitter vacua known from supergravity with compact
$R$-symmetry.

We also showed in Section 4 that for the four-dimensional case, these
solutions can be generated by a ``massive'' T-duality over the time,
which employs Scherk-Schwarz reductions to map the two
models. The model is similar to the type II$^{\star}$
models introduced by Hull \cite{120}.  In our case all kinetic terms
have the correct sign and this was possible due to a truncation of the
model on the bosonic sector.

Prototype examples of solutions of five- and four-dimensional
are of the form
\be990
\ba{lcr}
ds_5 &=& - e^{-4U} dt^2 + e^{2U} dx_4^2 \quad , \qquad
e^{6U} = \prod_{i=1}^3[\alpha_i \, t + h_i(x)] \\[2mm]
ds_4 &=& - e^{-2\hat U} dt^2 + e^{2\hat U} dx_3^2 \quad , \qquad
e^{4\hat U} = \prod_{i=1}^4[\alpha_i \, t + h_i(x)]
\ea \ee
where $h_i(x)$ are arbitrary harmonic functions of the four- or
three-dimensional flat transversal spatial space. If all constants
$\alpha_i$ vanish, we get back the well-known BPS solutions of
ungauged supergravity as, e.g.,\ the multi-center extremal black
holes. On the other hand, if these constants are non-zero these
multi-center black holes live in an space-time that asymptots at
late time to a de Sitter space with the cosmological constant
given by $\prod_i \alpha_i$.

These solutions are not supersymmetric, at least not in the usual
sense. However, they inherit properties well known from
supersymmetric solutions, e.g.,\ the existence of multi-center
solutions and the appearance of hyper-K\"ahler spaces.  Hence, we
expect that this de Sitter solution is stable, especially because
this the solution satisfies first order equations that coincide
with the BPS equations for a non-compact $R$-symmetry. The
existence of multi-center solutions implies a balance of forces,
namely the de Sitter expansion compensates the gravitational
attraction of the black holes.

Moreover, in the asymptotic de Sitter space the solution is in a
minimum of the potential with no tachyonic directions. The
potential coincides, up to the overall sign, with the potential
appearing in gauged supergravity with compact $R$-symmetry and
this potential has a number of interesting properties \cite{670}.
First of all, there is only one extremum of the superpotential on
a given component of the scalar manifold and it corresponds to a
maximum of the potential. For the model at hand the potential thus
has a stable minimum. Also, there are no flat directions, due to
the fact, that in the vacuum all  scalars of the
vector-supermultiplets are fixed. However, the situation changes
if one takes into account also scalar fields of
hyper-supermultiplets, which cannot be fixed completely in gauged
supergravity.

In general, there are no static coordinates to describe these
backgrounds.  One can introduce static coordinates only around a given
center or for the single-center solution \cite{590}.  Otherwise, this
solution is intrinsically time-dependent and if we naively continue to
early times, it exhibits a big bang singularity where the spatial part
of the metric collapses to a point.  Depending on the choice for the
harmonic function, the qualitative behavior near the singularity
differs. If e.g.\ in eq.\ (\ref{990}) $e^{4 \hat U}$ or $e^{6U}$
vanishes quadratically (setting $h_i(x) = const.$) the solution
exhibits a singularity as in the Milne universe, otherwise the
cosmological expansion near the singularity is decelerating for a
single zero or accelerating\footnote{We would like to thank Miguel S.\
Costa for a discussion on this point.}.  Note also, there is no reason
to stick to a time $t\geq 0 $ in the solutions above, and the
singularity is a repulson-type.  If this solution can be understood as
coming from a compactification, the scalar fields cannot naively be
extended to negative values.  In fact, the solution may run towards a
phase transition point which has to be treated with care. In
Subsection 3.3 we argued that, due to $\alpha'$ corrections, the
solution can run towards a regular de Sitter space at early time.

A different way of dealing with the singularity is to cut-off the
region by introducing a space-like brane at $t=0$, which is done
by replacing: $t \rightarrow |t|$. This however is a subtle issue,
because such a replacement requires a brane source. However, one
should allow for variations of the location of the brane source
and is seems that this will necessarily move the brane towards the
singularity, or at least  to the point where it becomes
tensionless. Although the solution remains basically the same,
these s-branes may provide a physical picture of the endpoint of
the flow. [Note, the s-branes appearing in our context are closely
related to the ones discussed in \cite{560,120}, however,  they
are of different nature as the ones discussed in \cite{310,333}
and references therein. In addition, the de Sitter vacua appearing
in our context are of different nature as the ones discussed for
non-compact gaugings as,  e.g., in \cite{740}.]

We mentioned also the possibility that the flow becomes completely
regular if one considers more general models with hyper-multiplets
where the superpotential has two continuously connected extrema
\cite{620, 630}. These potentials would describe a cosmological
scenario with an early and late time de Sitter vacuum, where,
depending on the nature of the fixed point, the space-time is
exponentially large or small. However, one should investigate
whether for this case the truncation to the neutral sector can
consistently been done.

Moreover, the  examples that we presented here are  closely
related to BPS solutions and it would be interesting whether the
non-extreme solutions  as,  e.g.,\ discussed in \cite{500}, have
also a cosmological analog. Moreover, for the four-dimensional
model we considered only the models with diagonal metrics, which
may be of most interest. However, it remains to be shown that the
general stationary BPS solutions can also be generalized to
cosmological backgrounds.

\vspace{10mm}

\noindent
{\bf Acknowledgments}

\medskip

\noindent
We would like to
thank the New Center for Theoretical Physics at Rutgers University and
the Institute for Advanced Study, Princeton, for hospitality and
support during the course of this work.  The research of M.C. is
supported in part by DOE grant DOE-EY-76-02-3071, NATO the linkage
grant No. 97061, NSF grant No. INT02-03585, and the Fay R. and Eugene
L. Langberg Chair.  The work of K.B.\ is supported by a Heisenberg
grant of the DFG.


\section*{Appendix}


In this appendix, we will show that the fields given in eqs.\
(\ref{100}) and (\ref{273}) solve in fact the first order differential
equations (\ref{625}).  Basically, we will repeat the calculations
done in \cite{590}, but we use a different notation and in order to
compare both results one has to rescale the harmonic functions
combined with a reparameterization of the time ($e^{-gt} \rightarrow g
t$).

Since we know the BPS solution as given in eq.\ (\ref{100}) and
(\ref{727}) \cite{130}, it is sufficient to look at the terms,
which are proportional to time-like derivatives -- all other terms
will cancel, because our starting point was a BPS configuration.
Using the relation $X_I dX^I$, see (\ref{251}), and the harmonic
functions as given in eq.\ (\ref{273}) one finds
\be193
\partial_0 e^{2U} = X^I \partial_0(e^{2U} X_I) = \lambda \alpha_I X^I
= \lambda W \ .
\ee
This yields for the term containing the superpotential
\be528
\ba{l}
g \Gamma_0 W = g \, \Gamma_{\underline 0} \, e^{-2U} W =
 { g \over  \lambda} \, \dot U \, \Gamma_{\underline 0} \ , \\
g \Gamma_m \, W = (e^{-2U} \omega_m \Gamma_{\underline 0}
+ e^U \Gamma_{\underline m}) \, W =
(\omega_m \Gamma_{\underline 0}
+ e^{3U} \Gamma_{\underline m}) \, {1 \over \lambda} \dot U \, W
\ea
\ee
where we have underlined the tangent space indices.  Moreover, the
gauge field component gives
\be153
G_{IJ} F^I = - {1 \over 2} e^{-4U} [ d H_I \wedge (dt + \omega_m dx^m)
 -  H_I d \omega ]
\ee
and because $X^I d H_I = 3 d e^{2U}$ we find for the time-like component
(note factors of 2/3 in (\ref{251})!)
\be794
X_I F^I_{0m} = (\partial_0 e^{-2U}) e^{2U} X_I A^I_m - X_I \partial_m A_0^I
= - 2 \, \dot U e^{-2U} \omega_m - \partial_m e^{-2U} \ .
\ee
Using the inverse metric
\be639
g^{00} = - e^{4U} + e^{-2U} \omega_m \omega_m
\quad , \qquad g^{0n} = - e^{-2U} \omega_n
\quad , \qquad g^{mn} = e^{-2U} \delta_{mn}
\ee
we can also calculate the field strength with upper indices. The
fact that $\omega_m \partial_m H_I = \omega_m \partial_m U = 0$
means that $F^{I \, mn}$ does not contain any time derivatives and
that
\be666
G_{IJ} F^{J\, 0n} = e^{-2U} \Big[ \partial_n H_I
- \omega_n \partial_0 H_I - e^{-6U} H_I \omega_m \partial_{[n} \omega_{m]}
\Big]  \ .
\ee
With these expressions it is now straightforward to calculate
the expression appearing in (\ref{625}). We get
\be527
\ba{l}
X_I (\Gamma_{0mn} F^{I\, mn} - 4 F_{0n}^I \Gamma^n)
= 8 \, e^{-3U} \dot U \omega_n \Gamma^{\underline n} + ... \\
X_I(2 \, \Gamma_{mn0} F^{I\, n0} - 4 F_{m0}^I  \Gamma^0 ) =
-2 \, \dot U \Gamma_{\underline {mn0}} \, \omega^n -8 \, \dot U
\omega_m ( \Gamma^{\underline{0}} - e^{-3U} \omega_n \Gamma^{\underline{n}})
+ ...
\ea
\ee
(we dropped all terms that have no time derivatives).  Next, for the
covariant derivative: $\nabla_\mu \epsilon = (\partial_\mu + {1 \over
4} \omega^{mn}_\mu \Gamma_{\underline{mn}} )\, \epsilon$ we find the
modifications due to the time-dependence
\be110 \ba{rcl} \nabla_0 \, \epsilon &=& (\hat \nabla_0 + e^{-3U} \dot
U \omega_m \Gamma_{\underline 0} \Gamma^{\underline{m}} ) \, \epsilon
\ , \\ \nabla_m \, \epsilon &=& (\hat \nabla_m + \dot U [ e^{-3U}
\omega_m \omega_n - {1 \over 2} e^{3U} \delta_{mn} ]
\Gamma_{\underline {0}} \Gamma^{\underline {n}} - {1 \over 4} \dot U
\omega_n \Gamma_{\underline m}^{\ \underline n} ) \, \epsilon \ .  \ea
\ee
Inserting all terms into (\ref{625}) one gets finally the differential
equation for $\epsilon$
\be552
[\partial_\mu  + (\partial_\mu U) ] \, \epsilon  = 0
\ee
which gives $\epsilon = e^{-U} \epsilon_0$ with the constant spinor
$\epsilon_0$ fulfilling the projector equation
\be739
\Gamma_{\underline 0} \epsilon_0 = i \epsilon_0 \ .
\ee
As next step, we have to verify the second equation in
(\ref{625}). Inserting the field $X_I$ from eq.\ (\ref{100}) we find
\bea710
\partial_A X^I \Big[ \Gamma^\mu \partial_\mu X_I  \Big]
&=&\partial_A X^I  \Big[ (\Gamma^{\underline 0}-
e^{-3U} \omega_m \Gamma^{\underline m} )  {1 \over 3}
 \alpha_I \lambda  + ... \Big]  \\
\partial_A X^I \Big[ G_{IJ} F^{J\, \mu m} \Gamma_{\mu m}  \Big]
&=& \partial_A X^I \Big[-  e^{-3U}  \omega_m
\Gamma_{ \underline{0m}} {1 \over 3}
\alpha_I \lambda  + ... \Big]
\eea
and with the projector (\ref{739}) we have verified also this equation.




\begin{thebibliography}{10}

\bibitem{120}
C.~M. Hull, ``Timelike {T}-duality, de {Sitter} space, large {N} gauge theories
  and topological field theory,'' {\em JHEP} {\bf 07} (1998) 021,
\href{http://www.arXiv.org/abs/hep-th/9806146}{{\tt hep-th/9806146}}.

\bibitem{510}
A.~Batrachenko, M.~J. Duff, and J.~X. Lu, ``The membrane at the end of the (de
  {Sitter}) universe,''
\href{http://www.arXiv.org/abs/hep-th/0212186}{{\tt hep-th/0212186}}.

\bibitem{660}
S.~Ferrara, ``Spinors, superalgebras and the signature of space-time,''
\href{http://www.arXiv.org/abs/hep-th/0101123}{{\tt hep-th/0101123}}.

\bibitem{690}
K.~Pilch, P.~van Nieuwenhuizen, and M.~F. Sohnius, ``De {S}itter superalgebras
  and supergravity,'' {\em Commun. Math. Phys.} {\bf 98} (1985)
105.

\bibitem{600}
D.~Kastor and J.~Traschen,
``Cosmological multi - black hole solutions,''
Phys.\ Rev.\ D {\bf 47} (1993) 5370
{\tt arXiv:hep-th/9212035}.
L.~A.~J. London, ``Arbitrary dimensional cosmological multi - black holes,''
  {\em Nucl. Phys.} {\bf B434} (1995)
709--735.
T.~Shiromizu,
``Cosmological spinning multi-'black-hole' solution in string theory,''
Prog.\ Theor.\ Phys.\  {\bf 102} (1999) 1207
{\tt arXiv:hep-th/9910176}.


\bibitem{590}
J.~T. Liu and W.~A. Sabra, ``Multi-centered black holes in gauged {D} = 5
  supergravity,'' {\em Phys. Lett.} {\bf B498} (2001) 123--130,
\href{http://www.arXiv.org/abs/hep-th/0010025}{{\tt hep-th/0010025}}.
%
D.~Klemm and W.~A. Sabra, ``General (anti-)de {Sitter} black holes in five
  dimensions,'' {\em JHEP} {\bf 02} (2001) 031,
\href{http://arXiv.org/abs/hep-th/0011016}{{\tt hep-th/0011016}}.

\bibitem{CGR}
M.~Cveti{\v c}, S.~Griffies, and S.-J. Rey, ``Static domain walls in {N=1}
  supergravity,'' {\em Nucl. Phys.} {\bf B381} (1992) 301--328,
\href{http://www.arXiv.org/abs/hep-th/9201007}{{\tt hep-th/9201007}}.

\bibitem{CGS}
M.~Cveti{\v c}, S.~Griffies, and H.~H. Soleng, ``Local and global gravitational
  aspects of domain wall space-times,'' {\em Phys. Rev.} {\bf D48} (1993)
  2613--2634,
\href{http://www.arXiv.org/abs/gr-qc/9306005}{{\tt gr-qc/9306005}}.

\bibitem{CS}
M.~Cveti{\v c} and H.~H. Soleng, ``Supergravity domain walls,'' {\em Phys.
  Rept.} {\bf 282} (1997) 159--223,
\href{http://www.arXiv.org/abs/hep-th/9604090}{{\tt hep-th/9604090}}.


\bibitem{620}
D.~Z. Freedman, S.~S. Gubser, K.~Pilch, and N.~P. Warner, ``Renormalization
  group flows from holography supersymmetry and a c-theorem,'' {\em Adv. Theor.
  Math. Phys.} {\bf 3} (1999) 363--417,
\href{http://www.arXiv.org/abs/hep-th/9904017}{{\tt hep-th/9904017}}.

\bibitem{760}
J.~M. Maldacena, ``The large {N} limit of superconformal field theories and
  supergravity,'' {\em Adv. Theor. Math. Phys.} {\bf 2} (1998) 231--252,
\href{http://www.arXiv.org/abs/hep-th/9711200}{{\tt hep-th/9711200}}.
%
S.~S. Gubser, I.~R. Klebanov, and A.~M. Polyakov, ``Gauge theory correlators
  from non-critical string theory,'' {\em Phys. Lett.} {\bf B428} (1998)
  105--114,
\href{http://www.arXiv.org/abs/hep-th/9802109}{{\tt hep-th/9802109}}.
%
E.~Witten, ``Anti-de {Sitter} space and holography,'' {\em Adv. Theor. Math.
  Phys.} {\bf 2} (1998) 253--291,
\href{http://www.arXiv.org/abs/hep-th/9802150}{{\tt hep-th/9802150}}.

\bibitem{160}
M.~Gunaydin, G.~Sierra, and P.~K. Townsend, ``Gauging the d = 5
  {M}axwell-{E}instein supergravity theories: More on {J}ordan algebras,'' {\em
  Nucl. Phys.} {\bf B253} (1985)
573.

\bibitem{150}
M.~Cveti{\v c} and D.~Youm, ``General rotating five dimensional black holes of
  toroidally compactified heterotic string,'' {\em Nucl. Phys.} {\bf B476}
  (1996) 118--132,
\href{http://www.arXiv.org/abs/hep-th/9603100}{{\tt hep-th/9603100}}.
%
J.~C. Breckenridge, R.~C. Myers, A.~W. Peet, and C.~Vafa, ``D-branes and
  spinning black holes,'' {\em Phys. Lett.} {\bf B391} (1997) 93--98,
\href{http://www.arXiv.org/abs/hep-th/9602065}{{\tt hep-th/9602065}}.

\bibitem{130}
A.~H. Chamseddine and W.~A. Sabra, ``Metrics admitting {K}illing spinors in
  five dimensions,'' {\em Phys. Lett.} {\bf B426} (1998) 36--42,
\href{http://www.arXiv.org/abs/hep-th/9801161}{{\tt hep-th/9801161}}.

\bibitem{201}
G.~W. Gibbons and S.~W. Hawking, ``Gravitational multi - instantons,'' {\em
  Phys. Lett.} {\bf B78} (1978)
430.

\bibitem{100}
J.~P. Gauntlett, J.~B. Gutowski, C.~M. Hull, S.~Pakis, and H.~S. Reall, ``All
  supersymmetric solutions of minimal supergravity in five dimensions,''
\href{http://arXiv.org/abs/hep-th/0209114}{{\tt hep-th/0209114}}.

\bibitem{210}
K.~Behrndt, D.~L{\"u}st, and W.~A. Sabra, ``Stationary solutions of {N} = 2
  supergravity,'' {\em Nucl. Phys.} {\bf B510} (1998) 264--288,
\href{http://www.arXiv.org/abs/hep-th/9705169}{{\tt hep-th/9705169}}.

\bibitem{390}
S.~Ferrara, R.~Kallosh, and A.~Strominger, ``N=2 extremal black holes,'' {\em
  Phys. Rev.} {\bf D52} (1995) 5412--5416,
\href{http://www.arXiv.org/abs/hep-th/9508072}{{\tt hep-th/9508072}}.
%
S.~Ferrara and R.~Kallosh, ``Supersymmetry and attractors,'' {\em Phys. Rev.}
  {\bf D54} (1996) 1514--1524,
\href{http://www.arXiv.org/abs/hep-th/9602136}{{\tt hep-th/9602136}}.

\bibitem{290}
K.~Behrndt, D.~L{\"u}st, and W.~A. Sabra, ``Moving moduli, {Calabi-Yau} phase
  transitions and massless {BPS} configurations in type {II} superstrings,''
  {\em Phys. Lett.} {\bf B418} (1998) 303--311,
\href{http://www.arXiv.org/abs/hep-th/9708065}{{\tt hep-th/9708065}}.

\bibitem{790}
M.~Cveti{\v c} and D.~Youm, ``Dyonic {BPS} saturated black holes
of heterotic string
  on a six torus,'' {\em Phys. Rev.} {\bf D53} (1996) 584--588,
\href{http://www.arXiv.org/abs/hep-th/9507090}{{\tt hep-th/9507090}}.
K.~Behrndt, R.~Kallosh, J.~Rahmfeld, M.~Shmakova, and W.~K. Wong, ``{STU} black
  holes and string triality,'' {\em Phys. Rev.} {\bf D54} (1996) 6293--6301,
\href{http://www.arXiv.org/abs/hep-th/9608059}{{\tt hep-th/9608059}}.


\bibitem{720}
A.~Strominger, ``The {dS/CFT} correspondence,'' {\em JHEP} {\bf 10} (2001) 034,
\href{http://www.arXiv.org/abs/hep-th/0106113}{{\tt hep-th/0106113}}.
A.~Strominger, ``Inflation and the {dS/CFT} correspondence,'' {\em JHEP} {\bf
  11} (2001) 049,
\href{http://www.arXiv.org/abs/hep-th/0110087}{{\tt hep-th/0110087}}.
%
V.~Balasubramanian, J.~de~Boer, and D.~Minic, ``Mass, entropy and holography in
  asymptotically de {Sitter} spaces,'' {\em Phys. Rev.} {\bf D65} (2002)
  123508,
\href{http://www.arXiv.org/abs/hep-th/0110108}{{\tt hep-th/0110108}}.

\bibitem{800}
K.~Skenderis and P.~K. Townsend, ``Gravitational stability and
  renormalization-group flow,'' {\em Phys. Lett.} {\bf B468} (1999) 46--51,
\href{http://www.arXiv.org/abs/hep-th/9909070}{{\tt hep-th/9909070}}.

\bibitem{670}
R.~Kallosh and A.~D. Linde, ``Supersymmetry and the brane world,'' {\em JHEP}
  {\bf 02} (2000) 005,
\href{http://www.arXiv.org/abs/hep-th/0001071}{{\tt hep-th/0001071}}.
%
K.~Behrndt and M.~Cveti{\v c}, ``Anti-de sitter vacua of gauged supergravities
  with 8 supercharges,'' {\em Phys. Rev.} {\bf D61} (2000) 101901,
\href{http://www.arXiv.org/abs/hep-th/0001159}{{\tt hep-th/0001159}}.

\bibitem{580}
M.~J. Duff, H.~L\"u, and C.~N. Pope, ``Heterotic phase transitions
and
  singularities of the gauge dyonic string,'' {\em Phys. Lett.} {\bf B378}
  (1996) 101--106,
\href{http://www.arXiv.org/abs/hep-th/9603037}{{\tt hep-th/9603037}}.

\bibitem{180}
M.~Cveti{\v c}, H.~L\"u, and C.~N. Pope, ``Brane resolution
through
  transgression,'' {\em Nucl. Phys.} {\bf B600} (2001) 103--132,
\href{http://www.arXiv.org/abs/hep-th/0011023}{{\tt hep-th/0011023}}.

\bibitem{300}
K.~Behrndt, ``About a class of exact string backgrounds,'' {\em Nucl. Phys.}
  {\bf B455} (1995) 188--210,
\href{http://www.arXiv.org/abs/hep-th/9506106}{{\tt hep-th/9506106}}.
%
R.~Kallosh and A.~D. Linde, ``Exact supersymmetric massive and massless white
  holes,'' {\em Phys. Rev.} {\bf D52} (1995) 7137--7145,
\href{http://www.arXiv.org/abs/hep-th/9507022}{{\tt hep-th/9507022}}.
%
M.~Cveti{\v c} and D.~Youm, ``Singular {BPS} saturated states and enhanced
  symmetries of four-dimensional {N}=4 supersymmetric string vacua,'' {\em
  Phys. Lett.} {\bf B359} (1995) 87--92,
\href{http://www.arXiv.org/abs/hep-th/9507160}{{\tt hep-th/9507160}}.

\bibitem{340}
C.~V. Johnson, A.~W. Peet, and J.~Polchinski, ``Gauge theory and the excision
  of repulson singularities,'' {\em Phys. Rev.} {\bf D61} (2000) 086001,
\href{http://www.arXiv.org/abs/hep-th/9911161}{{\tt hep-th/9911161}}.

\bibitem{370}
R.~Kallosh, T.~Mohaupt, and M.~Shmakova, ``Excision of singularities by stringy
  domain walls,'' {\em J. Math. Phys.} {\bf 42} (2001) 3071--3081,
\href{http://www.arXiv.org/abs/hep-th/0010271}{{\tt hep-th/0010271}}.
%
T.~Mohaupt, ``Topological transitions and enhancon-like geometries in
  {Calabi-Yau} compactifications of {M-theory},''
\href{http://www.arXiv.org/abs/hep-th/0212200}{{\tt hep-th/0212200}}.

\bibitem{350}
J.~Louis, T.~Mohaupt, and M.~Zagermann, ``Effective actions near
  singularities,'' {\em JHEP} {\bf 02} (2003) 053,
\href{http://www.arXiv.org/abs/hep-th/0301125}{{\tt hep-th/0301125}}.

\bibitem{BCN}
F.~A.~Brito, M.~Cveti\v c and A.~Naqvi, ``Brane Resolution And
Gravitational Chern-Simons Terms,'' Class.\ Quant.\ Grav.\  {\bf
20} (2003) 285. arXiv:hep-th/0206180.

\bibitem{480}
S.~B. Giddings, ``The fate of four dimensions,''
\href{http://www.arXiv.org/abs/hep-th/0303031}{{\tt hep-th/0303031}}.

\bibitem{490}
A.~Chou {\em et al.}, ``Critical points and phase transitions in 5d
  compactifications of {M-theory},'' {\em Nucl. Phys.} {\bf B508} (1997)
  147--180,
\href{http://www.arXiv.org/abs/hep-th/9704142}{{\tt hep-th/9704142}}.

\bibitem{570}
B.~R. Greene, K.~Schalm, and G.~Shiu, ``Dynamical topology change in {M}
  theory,'' {\em J. Math. Phys.} {\bf 42} (2001) 3171--3187,
\href{http://www.arXiv.org/abs/hep-th/0010207}{{\tt hep-th/0010207}}.

\bibitem{640}
A.~Ceresole, G.~Dall'Agata, R.~Kallosh, and A.~Van~Proeyen, ``Hypermultiplets,
  domain walls and supersymmetric attractors,'' {\em Phys. Rev.} {\bf D64}
  (2001) 104006,
\href{http://www.arXiv.org/abs/hep-th/0104056}{{\tt hep-th/0104056}}.

\bibitem{630}
K.~Behrndt and G.~Dall'Agata, ``Vacua of {N} = 2 gauged supergravity derived
  from non- homogenous quaternionic spaces,'' {\em Nucl. Phys.} {\bf B627}
  (2002) 357--380,
\href{http://www.arXiv.org/abs/hep-th/0112136}{{\tt hep-th/0112136}}.

\bibitem{250}
E.~Bergshoeff, M.~de~Roo, M.~B. Green, G.~Papadopoulos, and P.~K. Townsend,
  ``Duality of type {II} 7-branes and 8-branes,'' {\em Nucl. Phys.} {\bf B470}
  (1996) 113--135,
\href{http://www.arXiv.org/abs/hep-th/9601150}{{\tt hep-th/9601150}}.

\bibitem{410}
S.~Cecotti, S.~Ferrara, and L.~Girardello, ``Geometry of type {II} superstrings
  and the moduli of superconformal field theories,'' {\em Int. J. Mod. Phys.}
  {\bf A4} (1989)
2475.

\bibitem{420}
S.~Ferrara and S.~Sabharwal, ``Quaternionic manifolds for type {II} superstring
  vacua of {C}alabi-{Y}au spaces,'' {\em Nucl. Phys.} {\bf B332} (1990)
317.

\bibitem{430}
K.~Behrndt, I.~Gaida, D.~L{\"u}st, S.~Mahapatra, and T.~Mohaupt, ``From type
  {IIA} black holes to {T-dual} type {IIB} {D}-instantons in {N} = 2, {D} = 4
  supergravity,'' {\em Nucl. Phys.} {\bf B508} (1997) 659--699,
\href{http://www.arXiv.org/abs/hep-th/9706096}{{\tt hep-th/9706096}}.

\bibitem{440}
G.~W. Gibbons, M.~B. Green, and M.~J. Perry, ``Instantons and seven-branes in
  type {IIB} superstring theory,'' {\em Phys. Lett.} {\bf B370} (1996) 37--44,
\href{http://www.arXiv.org/abs/hep-th/9511080}{{\tt hep-th/9511080}}.

\bibitem{450}
J.~Scherk and J.~H. Schwarz, ``How to get masses from extra dimensions,'' {\em
  Nucl. Phys.} {\bf B153} (1979)
61--88.

\bibitem{110}
E.~Bergshoeff, M.~de~Roo, and E.~Eyras, ``Gauged supergravity from dimensional
  reduction,'' {\em Phys. Lett.} {\bf B413} (1997) 70--78,
\href{http://www.arXiv.org/abs/hep-th/9707130}{{\tt hep-th/9707130}}.

\bibitem{460}
I.~V. Lavrinenko, H.~L\"u, and C.~N. Pope, ``Fibre bundles and
generalised
  dimensional reductions,'' {\em Class. Quant. Grav.} {\bf 15} (1998)
  2239--2256,
\href{http://www.arXiv.org/abs/hep-th/9710243}{{\tt hep-th/9710243}}.
%
K.~Behrndt, E.~Bergshoeff, D.~Roest, and P.~Sundell, ``Massive dualities in six
  dimensions,'' {\em Class. Quant. Grav.} {\bf 19} (2002) 2171--2200,
\href{http://www.arXiv.org/abs/hep-th/0112071}{{\tt hep-th/0112071}}.

\bibitem{470}
E.~Bergshoeff and K.~Behrndt, ``D-instantons and asymptotic geometries,'' {\em
  Class. Quant. Grav.} {\bf 15} (1998) 1801--1813,
\href{http://www.arXiv.org/abs/hep-th/9803090}{{\tt hep-th/9803090}}.

\bibitem{520}
M.~Gutperle and M.~Spalinski, ``Supergravity instantons for {N} = 2
  hypermultiplets,'' {\em Nucl. Phys.} {\bf B598} (2001) 509--529,
\href{http://www.arXiv.org/abs/hep-th/0010192}{{\tt hep-th/0010192}}.

\bibitem{530}
J.~Louis and A.~Micu, ``Type {II} theories compactified on {Calabi-Yau}
  threefolds in the presence of background fluxes,'' {\em Nucl. Phys.} {\bf
  B635} (2002) 395--431,
\href{http://www.arXiv.org/abs/hep-th/0202168}{{\tt hep-th/0202168}}.


\bibitem{560}
C.~M. Hull and R.~R. Khuri, ``Branes, times and dualities,'' {\em Nucl. Phys.}
  {\bf B536} (1998) 219--244,
\href{http://www.arXiv.org/abs/hep-th/9808069}{{\tt hep-th/9808069}}.

\bibitem{310}
M.~Gutperle and A.~Strominger,
JHEP {\bf 0204} (2002) 018
[arXiv:hep-th/0202210].
\bibitem{333}
K.~Behrndt and S.~F{\"o}rste, ``String {Kaluza-Klein} cosmology,'' {\em Nucl.
  Phys.} {\bf B430} (1994) 441--459,
\href{http://www.arXiv.org/abs/hep-th/9403179}{{\tt hep-th/9403179}}.

\bibitem{740}
C.~M. Hull, ``Domain wall and de {Sitter} solutions of gauged supergravity,''
  {\em JHEP} {\bf 11} (2001) 061,
\href{http://www.arXiv.org/abs/hep-th/0110048}{{\tt hep-th/0110048}}.
%
G.~W. Gibbons and C.~M. Hull, ``de {Sitter} space from warped supergravity
  solutions,''
\href{http://www.arXiv.org/abs/hep-th/0111072}{{\tt hep-th/0111072}}.

\bibitem{500}
K.~Behrndt, M.~Cveti{\v c}, and W.~A. Sabra, ``Non-extreme black holes of five
  dimensional {N} = 2 {AdS} supergravity,'' {\em Nucl. Phys.} {\bf B553} (1999)
  317--332,
\href{http://www.arXiv.org/abs/hep-th/9810227}{{\tt hep-th/9810227}}.


\end{thebibliography}

\providecommand{\href}[2]{#2}\begingroup\raggedright\endgroup


\end{document}